\documentclass{aa}  
\usepackage{graphicx}
\usepackage{txfonts}

\usepackage[breaklinks=true, colorlinks=true, citecolor=blue, linkcolor=blue,urlcolor=blue]{hyperref}
\defcitealias{Beasley2025}{paper I}

\begin{document} 

   \title{Revisiting the globular clusters of NGC1052-DF2\thanks{Based on observations made with ESO Telescopes at the La Silla Paranal Observatory under programmes ID 2101.B-5008(A) and 110.23P4.001.}}

   \subtitle{}
   \titlerunning{Revisiting the globular clusters of NGC1052-DF2}

   \author{K. Fahrion\inst{1}\fnmsep\inst{2}
            \and
            M. A. Beasley\inst{3}\fnmsep\inst{4}\fnmsep\inst{5}
            \and
            A. Gvozdenko\inst{6}\fnmsep
            \and
            S. Guerra Arencibia\inst{3}\fnmsep\inst{4}
            \and
            T. Jerabkova\inst{7}\fnmsep\inst{8}
            \and
            J. Fensch\inst{9}
            \and 
            E. Emsellem\inst{7}
   }

    \institute{        
            Department of Astrophysics, University of Vienna, T\"{u}rkenschanzstra{\ss}e 17, 1180 Wien, Austria\\
            \email{katja.fahrion@univie.ac.at}
        \and
        European Space Agency, European Space Research and Technology Centre, Keplerlaan 1, 2201 AZ Noordwijk, the Netherlands
        \and
            Instituto de Astrofísica de Canarias, Calle V\'{i}a L\'{a}ctea, E-38206 La Laguna, Spain           
        \and    
            Departamento de Astrof\'{i}sica, Universidad de La Laguna, E-38206 La Laguna, Spain
        \and 
            Centre for Astrophysics and Supercomputing, Swinburne University, John Street, Hawthorn VIC 3122, Australia
        \and
        Department of Physics, Centre for Extragalactic Astronomy, Durham University, South Road, Durham DH1 3LE, UK
         \and
            European Southern Observatory, Karl-Schwarzschild-Strasse 2, 85748 Garching bei M\"unchen.
            \and
            Department of Theoretical Physics and Astrophysics, Faculty of Science, Masaryk University, Kotl\'{a}\v{r}sk\'{a} 2, Brno 611 37, Czech Republic
        \and
        Univ. Lyon, ENS de Lyon, Univ. Lyon1, CNRS, Centre de Recherche Astrophysique de Lyon, UMR5574, 69007 Lyon, France.
             }
   \date{\today}
 
  \abstract  
  {The ultra-diffuse galaxy (UDG) NGC\,1052-DF2 has captured the interest of astronomers ever since the low velocity dispersion measured from ten globular clusters (GCs) suggested a low dark matter fraction. Also, its GC system was found to be unusually bright, with a GC luminosity function peak at least one magnitude brighter than expected for a galaxy at a distance of 20 Mpc. In this work we present an updated view of the GC system of NGC\,1052-DF2. We analysed archival MUSE data of NGC\,1052-DF2 to confirm the membership of four additional GCs based on their radial velocities, thereby raising the number of spectroscopically confirmed GCs to 16. We measured the ages and metallicities of 11 individual GCs, finding them to be old ($> 9$ Gyr) and with a range of metallicities from [M/H] = $-0.7$ to $-1.8$ dex. The majority of GCs are found to be more metal-poor than the host galaxy, with some metal-rich GCs sharing the metallicity of the host ([M/H] = $-$1.09$^{+0.09}_{-0.07}$ dex). The host galaxy shows a flat age and metallicity gradient out to 1 $R_\text{e}$.
  Using a distance measurement based on the internal GC velocity dispersions ($D = 16.2$ Mpc), we derived photometric GC masses and find that the peak of the GC mass function compares well with that of the Milky Way. From updated GC velocities, we estimated the GC system velocity dispersion of NGC\,1052-DF2 with a simple kinematic model and find $\sigma_\text{GCS} = 14.86^{+3.89}_{-2.83}$ km s$^{-1}$. However, this value is reduced to $\sigma_\text{GCS} = 8.63^{+2.88}_{-2.14}$ km s$^{-1}$ when the GC that has the highest relative velocity based on a low S/N spectrum is considered an interloper. 
  We discuss the possible origin of NGC\,1502-DF2, taking the lower distance, spread in GC metallicities,  flat stellar population profiles, and dynamical mass estimate into consideration.}
               
   \maketitle

\section{Introduction}
Ultra-diffuse galaxies (UDGs) are a subclass of low-surface-brightness (LSB) galaxies characterised by their low luminosities and surface brightness ($\mu_g >$ 24 mag arcsec$^{-2}$) and large effective radii ($R_\text{e}$ > 1.5 kpc; \citealt{vanDokkum2015}). While the first examples of such galaxies were found in the 1980s (e.g. \citealt{SandageBinggeli1984, Impey1988, Dalcanton1997, ImpeyBothun1997, Conselice2003}), UDGs have received a lot of attention in the last decade thanks to ever more sensitive telescopes and deep photometric surveys (e.g. \citealt{Janssens2017, Prole2019, Zaritsky2019, Iodice2020, Zaritsky2023, Janssens2024}). Consequently, thousands of UDGs are now known, although the sample with available spectroscopy is notably smaller (but see for example \citealt{FerreMateu2023, Iodice2023, Gannon2024, FerreMateu2025}).

One UDG in particular has sparked a lot of interest from the research community: NGC\,1052-DF2 (also catalogued as KKSG04, PGC3097693, and [KKS2000]04; \citealt{Karachentsev2000}). Using Keck Low Resolution Imaging Spectrometer (LRIS) and DEep Imaging Multi-Object Spectrograph (DEIMOS) spectra of ten luminous globular clusters (GCs), \cite{vanDokkum2018_DM, vanDokkum2018_GC98_revision} found a surprisingly small velocity dispersion, $< 10$ km s$^{-1}$, suggesting a very low or even negligible fraction of dark matter (DM) contributing to the total mass of this galaxy. 

At the same time, \textit{Hubble} Space Telescope (HST) photometry showed that NGC\,1052-DF2 has an extremely luminous GC system (GCS) with a GC luminosity function (GCLF) turnover magnitude at least one magnitude brighter than expected if the galaxy is placed at $\sim$ 20 Mpc, the distance of the NGC\,1052 group \citep{vanDokkum2018_GCs}. To reconcile both the lack of DM and the luminous GCS, \cite{Trujillo2019} proposed that a distance of about 13 Mpc. At this distance, NGC\,1052-DF2 would be a more ordinary LSB galaxy. This lower distance was disputed with newer and deeper HST data by \cite{Shen2021_TRGB_distance}, who measured a distance of $D = 22.1 \pm 1.2$ Mpc based on the tip of the red giant branch (TRGB; see also \citealt{vanDokkum2018_distance}). In \cite{Shen2023}, this distance was revised to D = 21.7 Mpc to account for an incorrect alignment of the HST data. Using the same deep data, \cite{Shen2021_GCLF} show that this larger distance places the peak of the GCLF about 1.5 magnitudes brighter than the universal value of $M_V = -7.5$ mag. To add to these puzzling results, \cite{vanDokkum2022_GCs} find that the spectroscopically confirmed GCs all have very similar $V - I$ colours. They suggest that this is in line with a proposed formation scenario for NGC\,1052-DF2 and the UDG NGC\,1502-DF4, in which both galaxies formed after the collision of gas-rich progenitor galaxies \citep{vanDokkum2022_bullet, Tang2025}. Using a theoretical model, \cite{TrujilloGomez2021} also propose that the luminous GCS could have formed at very high gas densities, possibly in a major merger.

Beyond the original spectroscopic sample of ten GCs, \cite{Trujillo2019} present additional GC candidates, one of which was later confirmed as a GC by \cite{Emsellem2019}. These authors used deep integral field spectroscopy (IFS) with the Multi Unit Spectroscopic Explorer (MUSE; \citealt{MUSE}) to analyse the stellar body of NGC\,1052-DF2 and its GCS, confirming the low GCS velocity dispersion and finding a low velocity dispersion of the stars as well as indications of low-level stellar rotation. This small velocity dispersion of the stellar body was also found with  Keck Cosmic Web Imager (KCWI) IFS by \cite{Danieli2019}, who did not detect rotation in their smaller field of view. 

 \cite{Fensch2019} present an analysis of the stellar population properties of the host galaxy and the GCS based on the MUSE data. From an integrated spectrum of the stellar component and a stacked spectrum of the GCs within the MUSE field of view, they find old ages ($\sim 9$ Gyr). The stars appear to be more metal-rich ([M/H] = $-1.07 \pm 0.12$ dex) than the GCs ([M/H] $= -1.63 \pm 0.09$ dex). 
Similar values were found using slit spectroscopy by \cite{RuizLara2019}. From stacked Keck LRIS spectra, \cite{vanDokkum2018_GCs} also find an old age ($> 9$ Gyr) for the GCs but a higher average metallicity ([Fe/H] = $-1.35 \pm 0.12$ dex).

We re-analysed the MUSE data to confirm four additional GC candidates and to determine the ages and metallicities of individual GCs rather than stacked spectra. Complementary, an accompanying paper (\citealt{Beasley2025}, hereafter Paper I) presents the analysis of high-resolution FLAMES spectra. From measurements of the intrinsic GC velocity dispersion, Paper I presents an independent measurement of the distance to NGC\,1052-DF2 using the GC velocity dispersion distance method \citep{Beasley2024}, finding 16.2 $\pm$ 1.3 (stat.) $\pm$ 1.1 (sys.) Mpc. This method employs the relation between GC absolute magnitude and internal velocity dispersion to determine the distance. As described in 
\cite{Beasley2024}, the relation was calibrated using GCs of the Milky Way and M\,31 and was successfully applied to Centaurus\,A and Local Group dwarfs using literature data. In Paper I this method was applied to NGC\,1502-DF2, a much more distant galaxy. The found distance is at odds with the TRGB measurement, and as discussed in Paper I no definitive reason for this discrepancy is found. Paper I employs the metallicity measurements presented here and notes that intrinsic properties of the GCs (e.g. their mass-to-light ratios as derived from the MUSE spectra) agree with those of Milky Way GCs. A vastly different initial mass function might influence the results but only in combination with younger cluster ages, which are are disfavoured by the stellar population results presented here. Nonetheless, other as yet undiscovered systematic effects might influence this distance measurement. For consistency with Paper I, we assume a distance of 16.2 Mpc in the following but also discuss any distance-dependent results using the TRGB distance of 21.7 Mpc.
Combining the FLAMES results with the  analysis of the MUSE data described here, we present an updated view of NGC\,1052-DF2 and its GCS.

\begin{figure}
    \centering
    \includegraphics[width=0.47\textwidth]{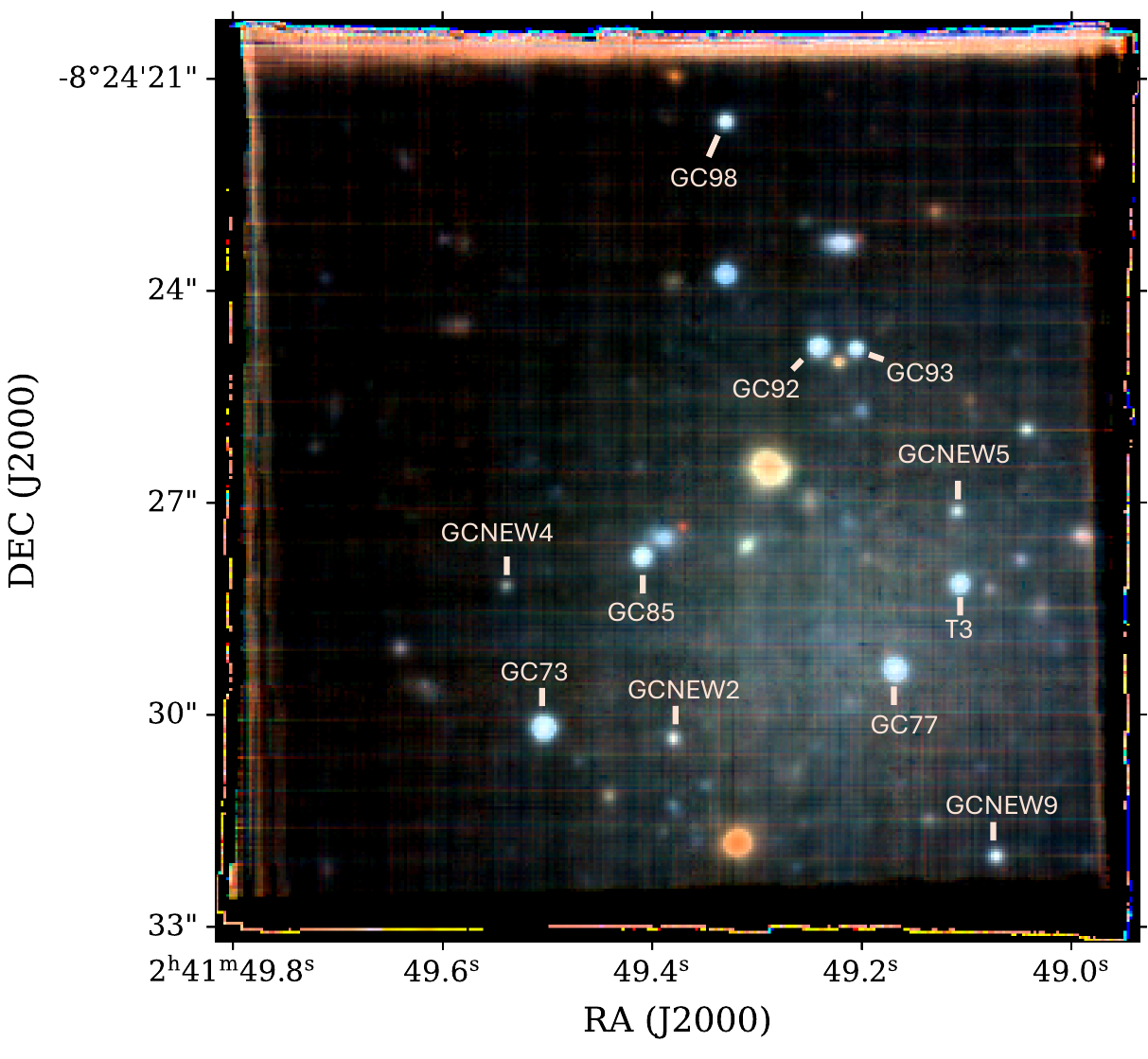}
    \caption{RGB image of NGC\,1052-DF2 created from the MUSE data. GCs analysed in this work are marked and labelled. Red: $V$, green: $R$, blue: $I$. See also \cite{Fensch2019}.}
    \label{fig:RGB_image}
\end{figure}

\section{Data}
\label{sect:data}

We used the fully reduced MUSE data as provided by \cite{Emsellem2019}. 
The data consist of seven observing blocks amounting to 5.1 h exposure time on target (programme 2101.B-5008(A)). The data were reduced following standard recipes including bias, flat fielding, wavelength calibration, and line spread function corrections using the dedicated python package \textsc{pymusepipe}\footnote{\url{https://github.com/emsellem/pymusepipe}}. Sky lines were removed using dedicated sky observations and residual sky lines were further cleaned by using the Zurich Atmosphere Purge (ZAP) code \citep{Soto2016}. \cite{Emsellem2019} provide different versions of the reduced data using different sky subtraction parameters. In this work, we used the linearly sampled, sky-subtracted version that was constructed using a 50 pixel continuum window in ZAP. 

The data cube covers a $\sim$1\arcmin $\times$ 1\arcmin\,field of view, sampled at 0.2\arcsec\,pix$^{-1}$ and with a point spread function (PSF) full width at half maximum (FWHM) of $\sim$ 0.8\arcsec. Along the wavelength direction, the cube covers the optical wavelength range from $\sim$ 4700 to 9300\,\AA\,with a sampling of 1.25\,\AA\,per pixel. As described in \cite{Emsellem2019}, the line spread function varies from FWHM $\sim$ 2.9\,\AA\,at 5000 \AA\, to $\sim$ 2.5 \AA\,at 9000 \AA\, but is well described by the UDF10 profile of \cite{Guerou2017}, but not exactly the same as discussed in the appendix of \cite{Emsellem2019}.
Figure \ref{fig:RGB_image} shows a false-colour red-green-blue (RGB) image obtained from the MUSE data using \textsc{mpdaf} \citep{MPDAF} to obtain synthetic images in $V$, $R$, and $I$. The GCs analysed in this work are marked and labelled.

As detailed in Paper I, several GCs of NGC\,1052-DF2 were observed with the FLAMES spectrograph. The spectra were taken with the H875.7 (HR21) grating that covers the Calcium triplet region (8500 - 9000 \AA) and have a spectral resolution of 0.53 \AA\, as measured from sky lines. Paper I focused on deriving the GC velocity dispersion from the five brightest GCs, while we present the radial velocities of the full sample in this work. For details on the data reduction and analysis of the FLAMES spectra, we refer to Paper I.

\section{Spectroscopic analysis}
\label{sect:analysis}
In the following, we describe our analysis of the analysis of the MUSE data. Firstly, we detail the extraction of GC spectra from the data and then describe the full spectrum fitting methods used to extract GC properties.

\subsection{Extracting GC spectra from MUSE}
To identify GCs in the data, we first constructed a white light image from the MUSE cube by integrating along the spectral axis. Using the \textsc{DAOStarFinder} routine implemented in \textsc{photutils} \citep{photutils}, we then detected all point sources 2$\sigma$ above the background flux level as determined in a corner of the image. By cross-referencing those sources with previously confirmed GCs \citep{vanDokkum2018_GCs, Emsellem2019, RuizLara2019}, we obtained the pixel centroids of GC73, GC77, GC85, GC92, GC93, GC98, and GCNEW3 (T3 in \citealt{Emsellem2019}), as well as the GC candidates GCNEW2, GCNEW4, GCNEW5, and GCNEW6 \citep{Trujillo2019}. Additionally, we added the coordinates of other bright point sources to check for yet undetected GCs.

To extract the spectra from MUSE, we used a similar method as in \cite{Fahrion2020b} and \cite{Fahrion2022}. To boost the signal from the GCs, we first obtained a model with \textsc{mpdaf} of the wavelength-dependent PSF using a bright foreground star (the source originally named GCNEW7 in \citealt{Trujillo2019}). We used this PSF cube as weighting for each wavelength slice. This approach leads to a slightly improved signal-to-noise ratio (S/N) compared to using a single two-dimensional Gaussian. The improvement is between 5 and 10 \%. We also extracted a spectrum of the underlying galaxy in an annulus around each GC. After testing, we chose to not subtract this background spectrum from the GC spectra as the contrast between GCs and galaxy is large and subtraction would therefore only introduce noise. Nonetheless, the derived properties remain even when subtracting the background spectrum, but with larger uncertainties.
Except for the faint source GCNEW6 (S/N $\sim$ 5), the resulting spectra have signal-to-noise values between 10 and 50 per pixel. As shown in the appendix of \cite{Fahrion2019}, such a S/N is sufficient to measure stellar population properties in addition to the line-of-sight velocities. For this reason, we chose to derive ages and metallicities of individual GCs rather than using a stacked spectrum as done by \cite{Fensch2019} and \cite{vanDokkum2018_GCs}.

\begin{figure*}
    \centering
    \includegraphics[width=0.95\textwidth]{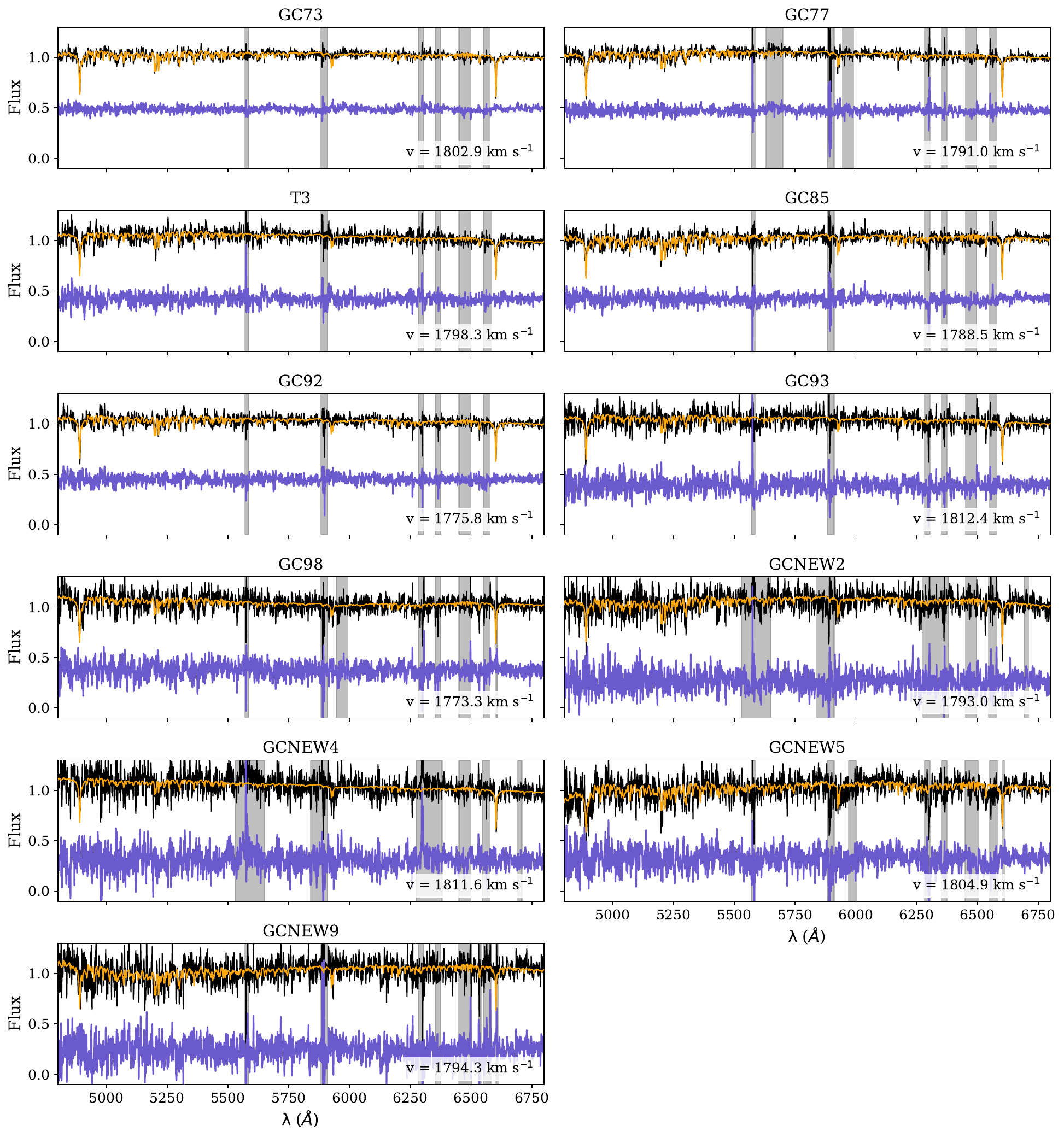}
    \caption{Full spectrum fitting of the GCs. Each panel shows the GC spectrum in black and the \textsc{pPXF} fit in orange. The respective residuals are shown in purple, shifted for visualisation. Grey shaded areas are regions masked from the fit due to sky residual or telluric lines. In the lower right corner, the best-fitting velocity is shown. The panels only show the blue part of the spectrum. }
    \label{fig:GC_spectra}
\end{figure*}

\begin{table*}[]
    \centering
    
    \caption{Spectroscopically confirmed GCs.}
    \begin{tabular}{c c c c c c c c c}\hline\hline
         ID & RA & Dec & $v_\text{MUSE}$ & $v_{\text{FLAMES}}$ & $v_{\text{Emsellem+2019}}$ & $v_{\text{van\,Dokkum+2018}}$ &  V$_{\text{F606W}}$ & I$_{\text{F814W}}$ \\
           & (J2000) & (J2000) & (km s$^{-1}$) & (km s$^{-1}$) & (km s$^{-1}$) & (km s$^{-1}$) & AB (mag) & AB (mag) \\
         (1) & (2) & (3) & (4) & (5) & (6) & (7) & (8)  & (9)  \\ \hline
GC39 & 40.43778 & $-$8.42357 & -- & $1800.6_{-1.8}^{+1.3}$ & -- & $1818.0_{-7.0}^{+7.0}$ & 22.37 $\pm$ 0.02 & 21.95 $\pm$ 0.02 \\
GC59 & 40.45037 & $-$8.41578 & -- & $1783.4_{-2.8}^{+3.2}$ & -- & $1799.0_{-15.0}^{+16.0}$ & 22.87 $\pm$ 0.03 & 22.31 $\pm$ 0.03 \\
GC71 & 40.43809 & $-$8.40620 & -- & $1794.4_{-3.7}^{+4.1}$ & -- & $1805.0_{-8.0}^{+6.0}$ & 22.65 $\pm$ 0.02 & 22.19 $\pm$ 0.03 \\
GC73 & 40.45093 & $-$8.40486 & $1802.9_{-3.7}^{+2.4}$ & $1797.5_{-1.3}^{+1.6}$ & $1803.8_{-3.0}^{+2.8}$ & $1814.0_{-3.0}^{+3.0}$ & 21.53 $\pm$ 0.02 & 21.06 $\pm$ 0.02 \\
GC77 & 40.44396 & $-$8.40371 & $1791.0_{-4.1}^{+4.2}$ & $1800.0_{-1.4}^{+1.0}$ & $1792.6_{-4.7}^{+4.4}$ & $1804.0_{-6.0}^{+6.0}$ & 22.10 $\pm$ 0.02 & 21.70 $\pm$ 0.02 \\
GC85 & 40.44896 & $-$8.40147 & $1788.5_{-3.8}^{+6.4}$ & $1792.1_{-0.9}^{+0.8}$ & $1786.3_{-4.8}^{+4.3}$ & $1801.0_{-6.0}^{+5.0}$ & 22.37 $\pm$ 0.02 & 21.84 $\pm$ 0.02 \\
GC91 & 40.42571 & $-$8.39813 & -- & $1796.7_{-3.3}^{+4.9}$ & -- & $1802.0_{-10.0}^{+10.0}$ & 22.51 $\pm$ 0.02 & 22.09 $\pm$ 0.03 \\
GC92 & 40.44536 & $-$8.39741 & $1775.8_{-5.9}^{+5.6}$ & $1780.2_{-1.0}^{+1.3}$ & $1775.5_{-4.2}^{+6.2}$ & $1789.0_{-7.0}^{+6.0}$ & 22.13 $\pm$ 0.02 & 21.63 $\pm$ 0.02 \\
GC93 & 40.44467 & $-$8.39743 & $1812.4_{-10.1}^{+6.6}$ & -- & $1819.1_{-6.8}^{+7.7}$ & -- & 22.82 $\pm$ 0.03 & 22.35 $\pm$ 0.03 \\
GC98 & 40.44726 & $-$8.39293 & $1773.3_{-8.4}^{+8.8}$ & $1777.4_{-5.2}^{+4.3}$ & $1786.3_{-11.6}^{+7.4}$ & $1784.0_{-10.0}^{+10.0}$ & 22.76 $\pm$ 0.02 & 22.29 $\pm$ 0.03 \\
GC101 & 40.43832 & $-$8.39104 & -- & $1836.0_{-2.6}^{+3.3}$ & -- & $1800.0_{-14.0}^{+13.0}$ & 22.86 $\pm$ 0.03 & 22.38 $\pm$ 0.03 \\
GCNEW2 & 40.44836 & $-$8.40506 & $1793.0_{-9.4}^{+9.8}$ & -- & -- & -- & 23.76 $\pm$ 0.04 & 23.25 $\pm$ 0.04 \\
GCNEW3 & 40.44264 & $-$8.40203 & $1798.3_{-7.7}^{+6.0}$ & -- & $1788.7_{-10.9}^{+16.9}$ & -- & 22.58 $\pm$ 0.02 & 22.16 $\pm$ 0.03 \\
GCNEW4 & 40.45166 & $-$8.40222 & $1811.6_{-15.4}^{+14.8}$ & -- & -- & -- & 24.08 $\pm$ 0.04 & 23.51 $\pm$ 0.05 \\
GCNEW5 & 40.44270 & $-$8.40079 & $1804.9_{-7.2}^{+9.4}$ & -- & -- & -- & 23.90 $\pm$ 0.04 & 23.43 $\pm$ 0.05 \\
GCNEW9 & 40.44205 & $-$8.40753 & $1794.3_{-10.7}^{+14.4}$ & -- & -- & -- & 23.70 $\pm$ 0.04 & 23.32 $\pm$ 0.04 \\
\hline
    \end{tabular}
         \tablefoot{(1): Cluster identifier as in \cite{vanDokkum2018_DM} and \cite{Trujillo2019}. (2), (3) right ascension and declination from HST data. (4) LOS velocity based on the here presented MUSE analysis, (5) velocities from the FLAMES spectra presented in Paper I, (6) velocities from MUSE as reported in \cite{Emsellem2019}, (7) velocities from Keck DEIMOS and LRIS spectra as reported in \cite{vanDokkum2018_DM, vanDokkum2018_GC98_revision}, (8), (9) apparent magnitudes from HST as detailed in Beasley et al. 2025}
    \label{tab:big_GC_table}
\end{table*}

\subsection{Full spectrum fitting with \textsc{pPXF}}
\label{sect:fitting}
We used the Penalized PiXel-Fitting \textsc{(pPXF)} method \citep{Cappellari2004, Cappellari2017, Cappellari2023} to fit the MUSE spectra. Using a penalised pixel fitting approach, \textsc{pPXF} fits a spectrum by finding the best-fitting linear combination of user-supplied models. \textsc{pPXF} allows the parameters of the line-of-sight velocity distribution to be measured, such as radial velocity and velocity dispersion. Additionally, \textsc{pPXF} returns the weights of the models used in the best fit and so the best-fitting age and metallicity can be reconstructed.

To fit the MUSE spectra, we used the EMILES single stellar population (SSP)\footnote{\url{http://research.iac.es/proyecto/miles/pages/spectral-energy-distributions-seds/e-miles.php}} models \citep{Vazdekis2010, Vazdekis2016} that are based on the MILES stellar spectra. The models are similar to those used by \cite{Emsellem2019} and \cite{Fensch2019}; however, we used the models based on the BaSTI isochrones \citep{Pietrinferni2004, Pietrinferni2006} with a double power-law initial mass function with high-mass slope of 1.30 \citep{Vazdekis1996}, rather than the Padova isochrones. The models cover ages from 30 Myr to 14 Gyr and total metallicities between [M/H] = $-2.27$ dex and $+0.40$ dex. 

During the fit, we used a two step approach: First, we fitted for the line-of-sight velocities using additive polynomials of degree 12. In the second step, the velocity is kept fixed and no additive polynomials are used. Instead, multiplicative polynomials of degree 8 are used to minimise template mismatch. From the returned weights, we inferred age and metallicity of the best-fitting spectrum. In the fit, sky residual lines are masked and a broad wavelength range from 4750 to 8750 \AA\, was used. Figure \ref{fig:GC_spectra} shows the blue part of the MUSE spectra and the \textsc{pPXF} fits.

To derive statistical uncertainties, we used a Monte Carlo (MC) approach. Here, we perturbed the best-fitting spectrum with randomly drawn noise from the residual (best fit subtracted from original spectrum) to create 100 representations of the spectrum. Those are fitted to obtain histograms of velocity, age, and metallicity. The reported values in Tables \ref{tab:big_GC_table} and \ref{tab:MUSE} refer to the 16th, 50th, and 84th percentiles of these distributions.

As described in Paper I, we also fitted the FLAMES spectra with \textsc{pPXF}. To get reliable velocity dispersion measurements, we used empirical stellar spectra that have to be brought to an arbitrary common velocity to be used as models. Here, however, we focus solely on the radial velocities and therefore used the high-resolution models from \cite{Coelho2014} with metallicities [Fe/H] $\leq$ $-1.0$ dex. 
The line-of-sight velocities are reported in Table \ref{tab:big_GC_table}. Figure 1 in Paper I shows two examples of the spectra.

\subsection{Confirming additional GC candidates}
Beyond the already confirmed GCs, we found four additional sources in the MUSE data that have velocities in agreement with the systemic velocity of NGC\,1052-DF2 ($v_\text{sys}$ = 1792.9$^{+1.4}_{-1.8}$ km s$^{-1}$, \citealt{Emsellem2019}). Those are GCNEW2, GCNEW4, and GCNEW5 from the candidates of \cite{Trujillo2019} as well as one source that has not been listed yet. We name this source GCNEW9, following the naming scheme in \cite{Trujillo2019}. The spectrum of GCNEW6 is rather noisy and while we find a velocity of $v = 1856$ km s$^{-1}$, the uncertainty is above 50 km s$^{-1}$. For this reason, we cannot confirm the nature of this source and do not include it in the following analysis.

From the FLAMES spectra, we could measure velocities of ten GCs, all previously confirmed as GCs. Unfortunately, the spectra of the GC candidates GCNEW1 and GCNEW8 are too noisy to obtain a velocity estimate as the uncertainties are above 20 km s$^{-1}$. In our sample, we also include GC101. For this GC, we found a velocity of 1836.0$^{+3.3}_{-2.6}$ km s$^{-1}$, which is the highest relative velocity of all GCs in the sample. As discussed below and in Paper I, this elevated velocity was found consistently when testing different fitting approaches, but due to the rather low S/N of 3, we cannot exclude that this velocity is affected by systematics or that at least the uncertainties are underestimated.

\section{Results}
With the four newly confirmed GCs, the sample of spectroscopically confirmed GCs of NGC\,1052-DF2 is now 16. We list their coordinates, velocities, and apparent magnitudes in Table \ref{tab:big_GC_table}. In addition, Table \ref{tab:MUSE} reports the stellar population results of the GCs within the MUSE data.

\subsection{Comparison of radial velocities}
\label{sect:velocities}
\begin{figure*}
    \centering
    \includegraphics[width=0.92\textwidth]{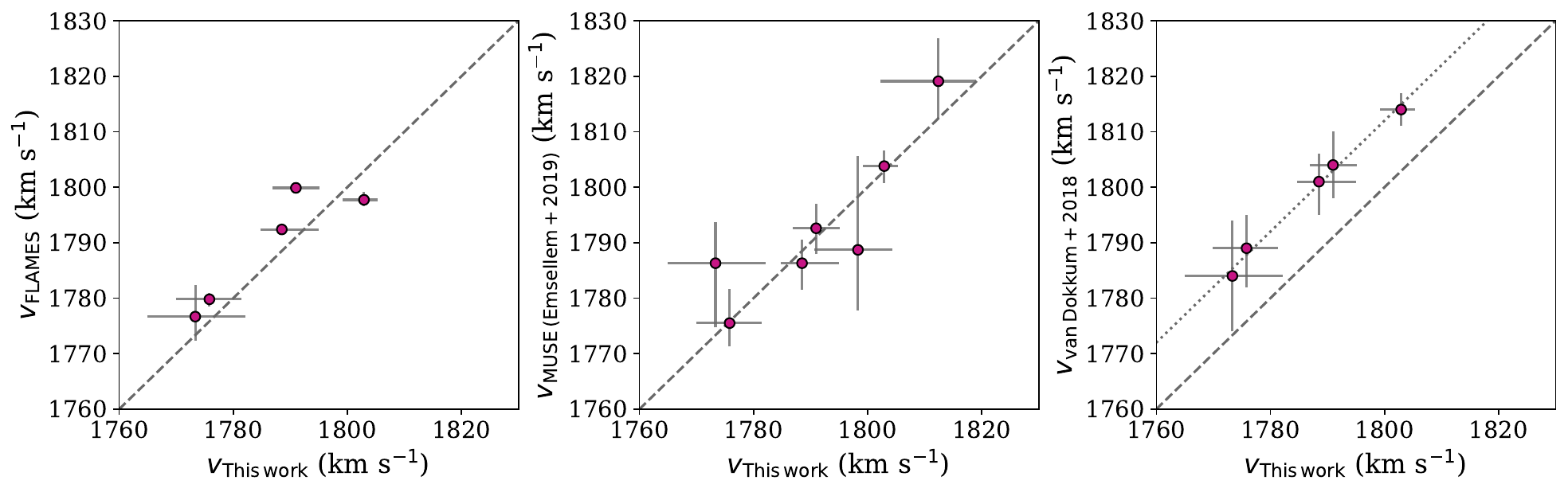}
    \caption{Comparison of velocity measurements. Left: Comparison of GC velocities from this work based on the MUSE data and from the FLAMES spectra presented in Paper I. Middle: Comparison between this work and GC velocities from the same data as presented in \cite{Emsellem2019}. Right: Comparison to GC velocities reported in \cite{vanDokkum2018_DM, vanDokkum2018_GC98_revision} from Keck DEIMOS and LRIS spectra. The dashed line shows the one-to-one relation, while the dotted line in the rightmost panel refers to a velocity offset of 12 km s$^{-1}$ (see also \citealt{Emsellem2019}).}
    \label{fig:velocity_comp}
\end{figure*}

We show a comparison of different GC velocity measurements in Fig. \ref{fig:velocity_comp}. In this figure, we compare our measurements from the MUSE spectra to the velocities obtained from the FLAMES spectra, as well as to the measurements from \cite{Emsellem2019} using the same MUSE data, and to the velocities from Keck DEIMOS and LRIS spectra \citep{vanDokkum2018_distance, vanDokkum2018_GC98_revision}. We find a good agreement between the velocities from MUSE and FLAMES; however, the measurements from FLAMES spectra have notably lower uncertainties. In addition, our measurements agree with those from \cite{Emsellem2019}, with a larger scatter for the lower S/N clusters, possibly due to a slightly different spectrum extraction method. As also noted in \cite{Emsellem2019}, there seems to be a consistent offset between the MUSE velocities and those from Keck DEIMOS and LRIS spectra of $\sim 12$ km s$^{-1}$. \cite{Emsellem2019} found an offset of 12.5$^{+2.6}_{-2.5}$ km s$^{-1}$ and suggested that part of it is caused by the transformation between redshift and wavelengths. Apart from this offset, the GC velocities between MUSE and DEIMOS/LRIS spectra agree very well.

\begin{table}[]
    \centering
    
    \caption{Globular clusters in the MUSE field of view.}
    \begin{tabular}{c c c c}\hline\hline
         ID  & [M/H] & Age & $M/L_\text{F606W}$ \\
         (dex) & (Gyr) & &  \\
         (1) & (2) & (3) & (4)  \\ \hline
GC73 & $-1.51_{-0.08}^{+0.10}$ & $9.6_{-1.3}^{+2.1}$ & 1.56 $\pm$ 0.27 \\
GC77 & $-1.61_{-0.09}^{+0.09}$ & $11.4_{-1.8}^{+0.8}$ & 1.41 $\pm$ 0.23 \\
GC85 & $-1.07_{-0.14}^{+0.13}$ & $9.2_{-2.0}^{+1.8}$ & 1.68 $\pm$ 0.26 \\
GC92 & $-1.53_{-0.10}^{+0.12}$ & $10.1_{-1.5}^{+2.1}$ & 1.45 $\pm$ 0.32 \\
GC93 & $-1.58_{-0.17}^{+0.13}$ & $10.6_{-2.2}^{+1.2}$ & 1.43 $\pm$ 0.33 \\
GC98 & $-1.70_{-0.12}^{+0.17}$ & $11.0_{-2.7}^{+1.1}$ & 1.35 $\pm$ 0.34 \\
GCNEW2 & $-1.18_{-0.20}^{+0.24}$ & $10.7_{-2.8}^{+1.9}$ & 1.88 $\pm$ 0.32 \\
GCNEW3 & $-1.78_{-0.15}^{+0.13}$ & $9.6_{-2.3}^{+2.5}$ & 1.16 $\pm$ 0.39 \\
GCNEW4 & $-1.86_{-0.25}^{+0.22}$ & $12.9_{-3.0}^{+1.1}$ & 1.25 $\pm$ 0.56 \\
GCNEW5 & $-0.73_{-0.17}^{+0.19}$ & $10.3_{-3.4}^{+2.3}$ & 2.04 $\pm$ 0.43 \\
GCNEW9 & $-0.92_{-0.24}^{+0.20}$ & $10.6_{-2.5}^{+2.1}$ & 1.98 $\pm$ 0.32 \\
\hline
    \end{tabular}
         \tablefoot{(1): Cluster identifier as in \cite{vanDokkum2018_DM} and \cite{Trujillo2019}. (2), (3) metallicities and ages the GCs, (4) mass-to-light ratio ($M/L$) in the F606W band from the stellar population fits.}
    \label{tab:MUSE}
\end{table}
\subsection{Age metallicity distribution}
\label{sect:age_metal_dist}
\begin{figure}
    \centering
    \includegraphics[width=0.48\textwidth]{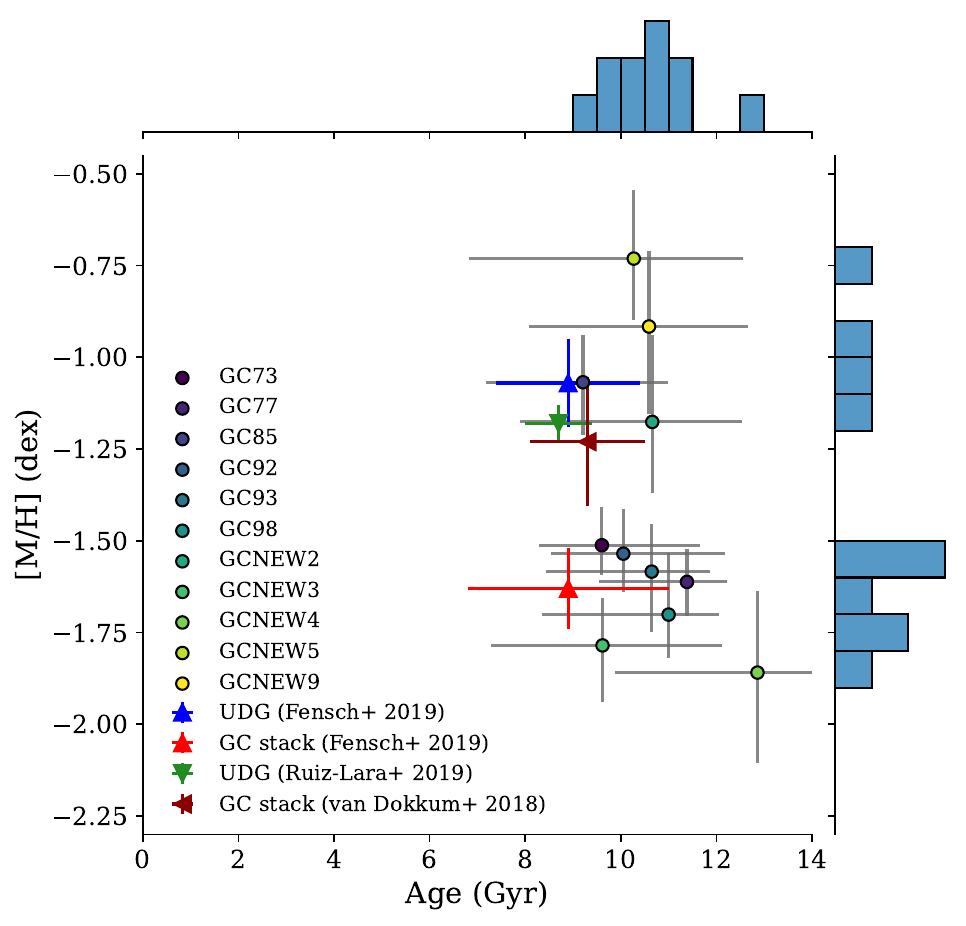}
    \caption{Age versus metallicity of the GCs. Coloured dots refer to the MUSE GCs analysed here. The red cross shows the result from the stacked GC spectrum of \cite{Fensch2019}, and the blue cross refers to the UDG itself as determined from an aperture spectrum. The dark red point shows the result from the stacked Keck spectra from \cite{vanDokkum2018_GCs}, and the green triangle shows the UDG metallicity from slit spectroscopy presented by \cite{RuizLara2019}.}
    \label{fig:age_metallicity}
\end{figure}

Figure \ref{fig:age_metallicity} shows the best-fitting ages and total metallicities of the analysed GCs, compared to results from \cite{vanDokkum2018_GCs}, \cite{Fensch2019}, and \cite{RuizLara2019} based on stacked spectra. Using slit spectroscopy, \cite{RuizLara2019} found an average age of 8.7 $\pm$ 0.7 Gyr and a metallicity of [M/H] = $-1.18 \pm 0.05$ dex, similar to the results from an aperture spectrum extracted from the MUSE data by \citep{Fensch2019}. Additionally, \cite{Fensch2019} used the MUSE spectra of the six GCs presented in \cite{Emsellem2019} to create a stacked spectrum, weighted by S/N, and fitted with \textsc{pPXF} and the EMILES models. \cite{vanDokkum2018_GCs} stacked LRIS-blue spectra of 11 GCs and fitted the spectrum with \textsc{alf} \citep{Conroy2012, Conroy2018} to obtain a best-fitting iron metallicity of [Fe/H] = $-1.35 \pm 0.12$ dex, magnesium abundance of [Mg/Fe] = $+$0.16 $\pm$ 0.17 dex, and age = 9.3$^{-1.3}_{-1.2}$ Gyr. To compare these iron metallicities with the total metallicities used in EMILES, we used the relation $[\text{Fe/H}] = [\text{M/H}] - 0.75 \times [\text{Mg/Fe}]$ as given on the MILES web page. 
As already noted by \cite{Fensch2019}, their GC metallicity is lower than what was found by \cite{vanDokkum2018_GCs} and they discussed multiple reasons for the difference in measured GC metallicity, including the different modelling approaches, different SSP models, and wavelength ranges. 

With our individual measurements and improved GC number, we can now investigate the age-metallicity plane in greater detail.
Overall, the GCs appear to be old with ages $>$ 9 Gyr, similar to what was found from the stacked spectra. However, we observe a broad range of GC metallicities, ranging from [M/H] = $-$0.73 dex to $-1.86$ dex. Out of the eleven GCs in the MUSE field of view, seven have metallicities $< -1.50$ dex, while the remaining four have higher metallicities. Interestingly, the metal-poor GCs appear to scatter around the average value found by \citep{Fensch2019}, while the more metal-rich ones have similar metallicities as the UDG itself.
Out of the GCs presented here, \cite{Fensch2019} included six metal-poor GCs (GC73, GC77, GC92, GC93, GC98, and T3) in their stack as well as the metal-rich GC85, which explains the low average metallicity. Nonetheless, due to the higher uncertainties of the individual GC metallicities, even the metal-rich ones agree with the stacked value within 3$\sigma$, with the exception of GCNEW5. 
\cite{vanDokkum2018_GCs} included also GCs outside the MUSE field of view for which no individual metallicities are available. However, the higher average metallicity could be an indication that their stacked spectrum has also included a higher number of metal-rich GCs.

\begin{figure*}
    \centering
    \includegraphics[width=0.95\textwidth]{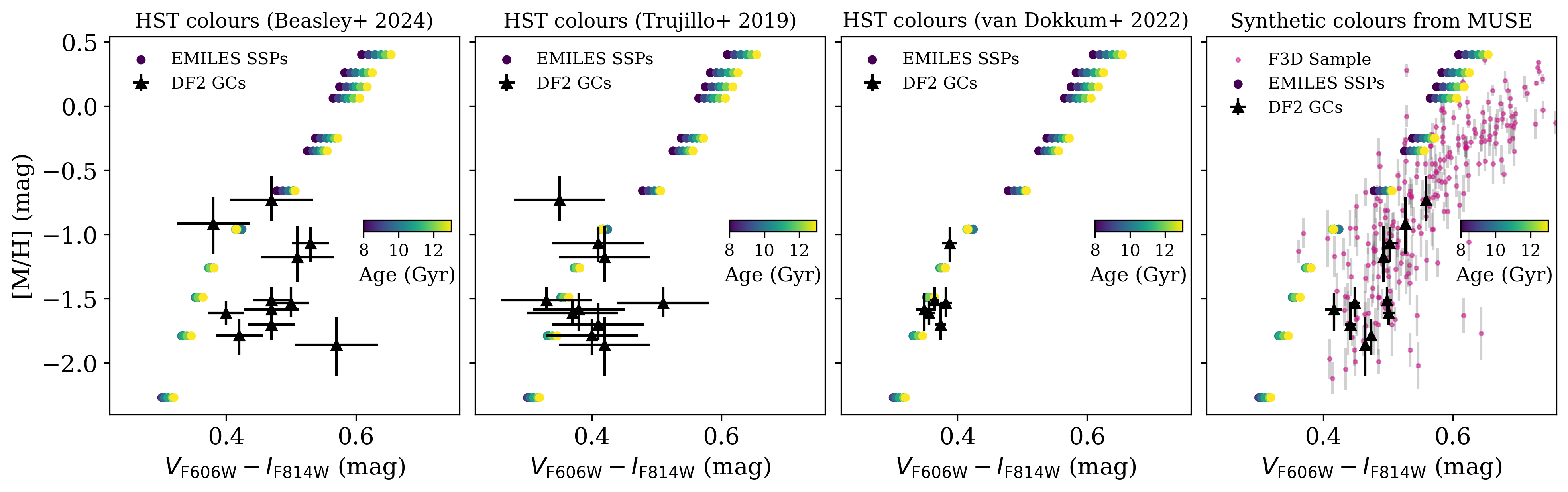}
    \caption{CZR of the GCs. Each panel shows the spectroscopic metallicities versus HST F606W $-$ F814W colour and predictions from the EMILES models colour-coded by age. From left to right: Colours from Paper I, \cite{Trujillo2019}, \cite{vanDokkum2022_GCs}, and synthetic colours derived from the MUSE spectra directly. In the right panel, we also include the Fornax3D GC sample from \cite{Fahrion2020b}.}
    \label{fig:CZR}
\end{figure*}

\subsection{Colour--metallicity relation}
\label{sect:CZR}
We investigated the relationship between photometric colours and spectroscopic metallicities in Fig. \ref{fig:CZR}. The slope of this so-called colour--metallicity relation (CZR) is important when photometric measurements are used as this relation allows one to translate from colours to the more fundamental property of metallicity (e.g. \citealt{Usher2015, Fahrion2020c, Hartmann2023}).

In Fig. \ref{fig:CZR} we show the GC metallicities versus $V_\text{F606W} - I_\text{F814W}$ colours in the left panel. The photometric measurements stem from a new analysis of the deeper HST data (combining data from 2017 and 2020), as described in more detail in Paper I. Additionally, we compare those colours to the HST measurements using the original HST data from \cite{Trujillo2019} as well as the colours from \cite{vanDokkum2022_GCs}. The latter are also based on the deeper HST data, but the colours from \cite{vanDokkum2022_GCs} are significantly bluer and span only a very narrow colour range. As also discussed in Paper I, we found F606W magnitudes that are $\sim$ 0.1 mag fainter than the ones reported in \cite{vanDokkum2022_GCs}, while the F814W magnitudes agree within our larger uncertainties. As a consequence, we found a larger spread in colours.
The reason for this discrepancy is unknown, but might be related to a different approach to modelling the photometric data. Unfortunately, the photometric measurements of \cite{vanDokkum2022_GCs} do not include the newly confirmed GCs that are more metal-rich, so the comparison to metallicities is restricted to only five GCs, four of which are metal-poor.

Additionally, we derived synthetic colours in the HST Advanced Camera for Surveys (ACS) F606W and F814W bands using \textsc{synphot} \citep{synphot} to convolve the MUSE spectra with the HST transmission curves. The same approach was taken to obtain colour predictions from the EMILES SSP models. As a comparison sample, we also obtained colours in the same bands from the spectra of the Fornax3D GC sample \citep{Fahrion2020b}, which provides a MUSE-based spectroscopic sample of 230 GCs with $S/N > 8$ around mainly massive early-type galaxies in the Fornax galaxy cluster. We note that the synthetic MUSE colours agree with the HST colours and compare well to the Fornax3D GC sample.

Irrespective of the differences in colours derived with different approaches, Fig. \ref{fig:CZR} clearly shows that a narrow range of colours does not necessarily imply a narrow range of metallicities. The reason for this is that at the blue colours of the NGC\,1052-DF2 GCs, the CZR has a steep slope, as for example the large Fornax3D sample illustrates. The same is seen for the EMILES SSP predictions that can give an indication for the slope of the CZR, even though the SSPs do not cover all the aspects of true GCs\footnote{We note that the mismatch between EMILES SSP predictions and observed colours does not affect the age and metallicity measurements because the \textsc{pPXF} fitting uses polynomials that make the result insensitive to the overall colour.}.

\subsection{GC mass-to-light ratios and masses from stellar populations}
\label{sect:MLs}Fitting the GC spectra for their ages and metallicities allows us to determine their mass-to-light ratios ($M/L$) as predicted from the EMILES SSP models. For this, we used the photometric predictions in the HST ACS filter F606W that provide the mass-to-light ratio for every age and metallicity in the model grid. We used a two-dimensional interpolation to obtain the $M/L$ for any given age and metallicity and used this to translate the MC distributions of age and metallicity to a $M/L$ value for every MC fit. From this, we determined the $M/L$ and its uncertainty. The results are reported in Table \ref{tab:big_GC_table}. The $M/L$ values range from 1.25 to 2.04, similar to typical $M/L$ values of GCs in the MW \citep{BaumgardtHilker2018} and also comparable to the dynamical $M/L$ values derived from the FLAMES spectra, which have a mean V-band $M/L$ = 1.61 $\pm$ 0.44 (Paper I).

\begin{figure}
    \centering
    \includegraphics[width=0.45\textwidth]{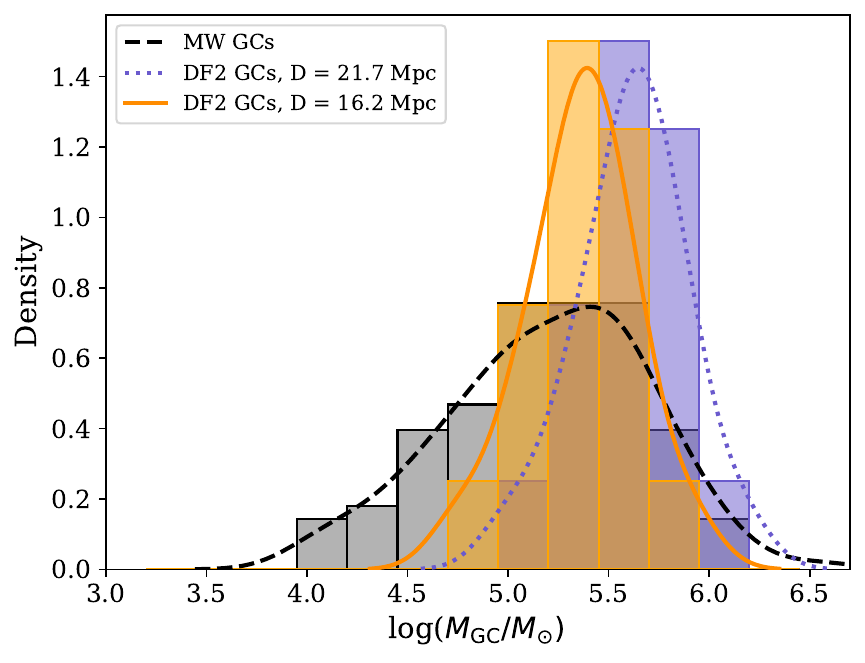}
    \caption{Distribution of derived masses based on stellar population properties. The orange histogram shows the NGC\,1052-DF2 GCs assuming a distance of 16.2 Mpc, and the orange line shows a kernel density estimation line to visualise the distribution. We note that the histogram only shows the spectroscopically confirmed GCs and that no completeness correction was applied. The purple histogram and corresponding dotted purple line shows the mass distribution assuming a larger distance of 21.7 Mpc. The grey histogram and the black line show the distribution of 112 GCs from the Milky Way from \cite{BaumgardtHilker2018}.}
    \label{fig:masses}
\end{figure}

To then estimate the stellar masses of the GCs, we used the GC apparent magnitudes from Paper I and assumed a distance of 16.2 Mpc (Paper I). For GCs outside the MUSE field of view without metallicities, we used the $M/L_{\text{F606W}}$ = 1.45, which corresponds to the median value. The resulting histogram of masses is shown in Fig. \ref{fig:masses}. We compare this histogram to Milky Way GCs from \cite{BaumgardtHilker2018} (black curve), showing that the peak that the mass distribution of the NGC\,1052-DF2 GCs analysed here is located at $\sim 10^{5.4} M_\sun = 2.5 \times 10^5 M_\sun$, very close to the canonical peak of the GC mass function at $M_\text{GC} \sim 2 \times 10^5 M_\odot$ \citep{Harris2001, Brodie2006, Jordan2007}. At a distance of 21.7 Mpc, the peak would be at higher masses of $\sim 4 \times 10^5 M_\sun$, as also noted by \cite{Shen2021_GCLF}.
We note that using the brighter F606W magnitudes presented in \cite{vanDokkum2022_GCs} also changes the mass distribution. For example, using their magnitudes and a distance of 16.2 Mpc, the peak of the distribution is then placed at 2.9 $\times$ 10$^{5}$ M$_\sun$.

\subsection{Comparison to the stellar light}
\label{sect:stellar_light}
In their analysis of the MUSE data, \cite{Fensch2019} reported an average metallicity and age of the stellar light obtained from fitting an aperture spectrum within 1 $R_\text{e}$ (22\arcsec).
Here, we provide more detail by extracting the galaxy spectrum in different annuli out to the effective radius. For this, we first masked GCs and other compact sources from the data cube (similar to \citealt{Emsellem2019}). Then, we extracted the galaxy spectrum in elliptical annuli with position angle = $-48^{\circ}$\, and ellipticity $\epsilon = 0.15$ \citep{vanDokkum2018_distance, Emsellem2019}. As annuli widths we chose 10 pix ($= 2\arcsec$) out to 70 pix and then chose 20 pix for the two outermost bins due to the decreasing surface brightness. We then fitted the spectra with \textsc{pPXF}.

\begin{figure}
    \centering
    \includegraphics[width=0.45\textwidth]{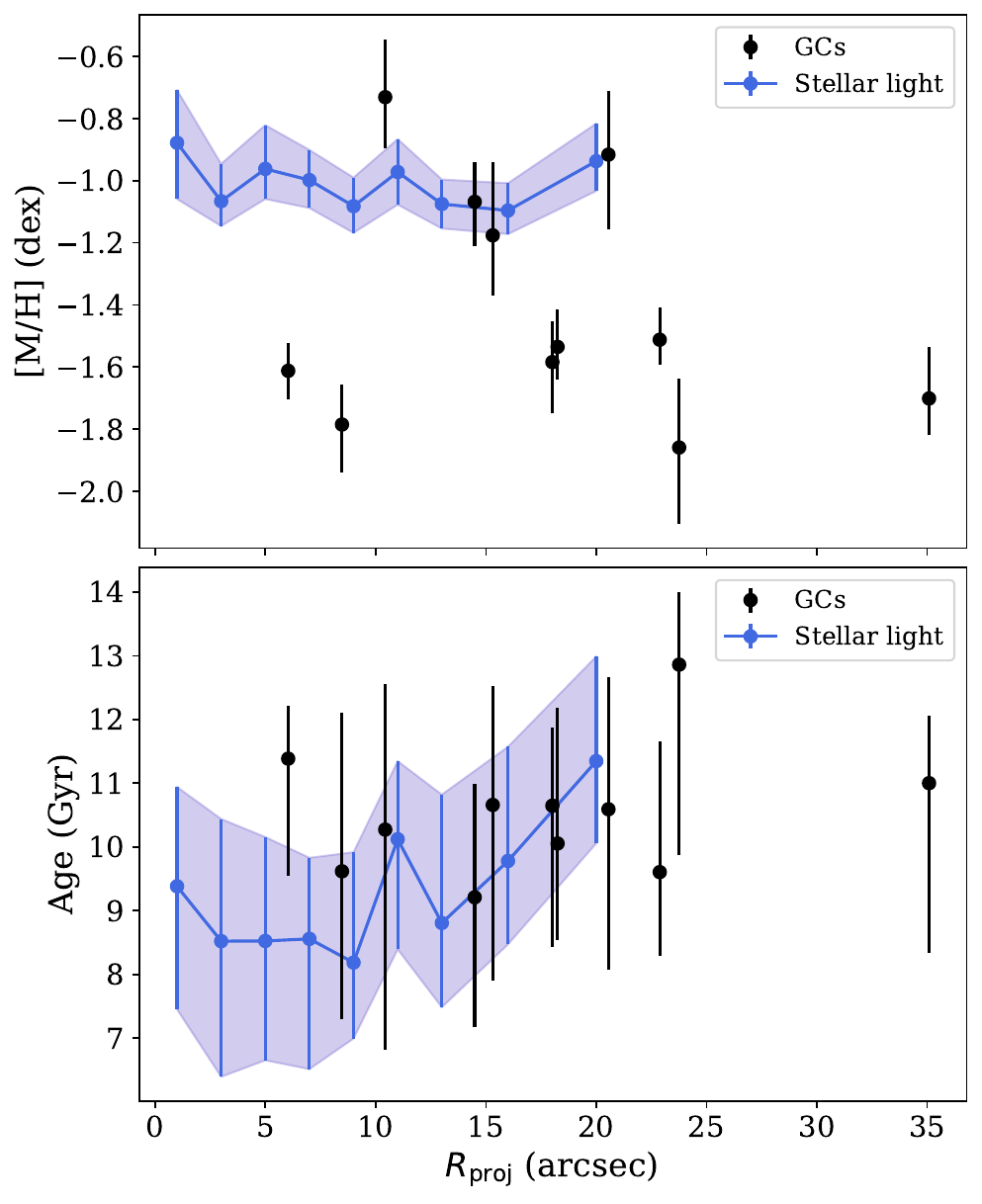}
    \caption{Radial metallicity and age profiles. Top: Metallicity profile of the stellar light obtained from elliptical annuli spectra in blue and GC metallicities in black. Bottom: Age profile for the stars and the GCs.}
    \label{fig:stellar_pop_profiles}
\end{figure}

Figure \ref{fig:stellar_pop_profiles} shows the radial profiles of metallicities and ages of the stellar light in comparison to the GCs. Within the uncertainties, both age and metallicity profiles appear to be flat, but there seems to be a mild trend of older ages at larger radii. While many low group dwarf galaxies are found to have negative metallicity gradients (e.g. \citealt{Taibi2022}), the UDG DF44 is also known to have a flat metallicity profile \citep{Villaume2019}. However, the sample of UDGs with spectroscopically measured stellar population gradients is very limited (see e.g. \citealt{Gannon2024, FerreMateu2025}).

As noted before, the GCs are overall old, but scatter in metallicity over a wide range, with most being more metal-poor than the stars. In a galaxy with an extended star formation history, such a spread is expected. The more metal-rich GCs scatter around the metallicity of the host galaxy, as also often observed for more massive galaxies (e.g. \citealt{Fahrion2020b}). 
In addition to the annuli spectra, we also obtained integrated spectra within the effective radius. From this, we found [M/H] = $-1.07^{+0.09}_{-0.07}$ dex and Age = $8.9^{+1.4}_{-1.5}$ Gyr, in agreement with the findings from \cite{Fensch2019}.  For these values, the EMILES models predict a $M/L_\text{F606W}$ of 1.67 $\pm 0.17$ in the HST F606W filter. With an apparent magnitude of $V = 16.1$ mag \citep{vanDokkum2018_DM}, this mass-to-light ratio translates to a stellar mass of $M_\ast \sim 1.1 \times 10^{8} M_\sun$ at 16.2 Mpc, or $M_\ast \sim 2 \times 10^8 M_\sun$ at 21.7 Mpc. 

\subsection{Simple kinematic model of the GC system}
\label{sect:kin_model}
Combing the samples from MUSE and FLAMES, we now have a sample of 16 GCs in NGC\,1052-DF that have precise velocity measurements available. For this reason, we model the kinematics of the GCS in a similar way as described in \cite{Martin2018}, \cite{Laporte2019}, \cite{Lewis2020_GCrotation}, or \cite{Emsellem2019}, but with an updated number of GCs. For our model, we used the same model as in \cite{Fahrion2020b}, which follows the description in \cite{Veljanoski2017} and is comparable to the rotation model by \cite{Lewis2020_GCrotation}.
In the model, the global rotation of the GCS is described as \citep{Cote2001}
\begin{equation}
    v_\text{GC,i} (\theta) = v_\text{sys} + V_\text{GCS}\,\text{sin}(\theta_i - \theta_0),
\end{equation}
where $v_\text{GC,i}$ is the velocity of the $i$th GC at a position angle of $\theta_i$, $v_\text{sys}$ the systemic velocity of the galaxy and $V_\text{GCS}$ the rotation amplitude. $\theta_0$ describes the angle of the rotation axis, where the rotation is maximum at $\theta_0$ + 90$^{\circ}$. In this model, only the global rotation and dispersion is described and it assumes that those do not vary with radius.

Assuming that the intrinsic velocity dispersion $\sigma_\text{GCS}$ is well represented by a Gaussian, it can be described as\begin{equation}
    \sigma^2_{\text{obs}} = (\delta v_\text{GC,i})^2 + \sigma_\text{GCS}^2,
\end{equation}
where $\delta v_\text{GC,i}$ is the velocity uncertainty of the $i$th GC. We then write the likelihood function as
\begin{equation}
\mathcal{L} = \prod \limits_{i} \frac{1}{\sqrt{2 \pi \sigma_\text{obs}^2}} \text{exp} \left(- \frac{(v_\text{GC, i} - (v_0 + V_\text{GCS}\,\text{sin}(\theta_i - \theta_\text{kin})))^2}{2 \sigma_\text{obs}^2} \right).
\label{eq:kin_model}
\end{equation}
As input to this model, we derived the position angles $\theta_i$ of the GCs from the north going counter-clockwise and assuming a galaxy centre at RA = 02:41:46.80, Dec = $-$08:24:09.3. For the velocities, we used the FLAMES velocities where available and the MUSE velocities for the rest. The uncertainties were converted to symmetric values by taking the mean between lower and upper uncertainty.

We implemented the model in \textsc{emcee} \citep{emcee}, a python module that implements a Markov chain Monte Carlo sampler. 
We used flat priors for all free parameters ($v_\textsc{sys}$, $V_\text{GCS}$, $\sigma$, $\theta_0$). We restricted the position angle to be between 0 and 180$^\circ$, but allowed the rotation amplitude to also take negative values to describe counter-rotation.

\begin{figure}
    \centering
    \includegraphics[width=0.47\textwidth]{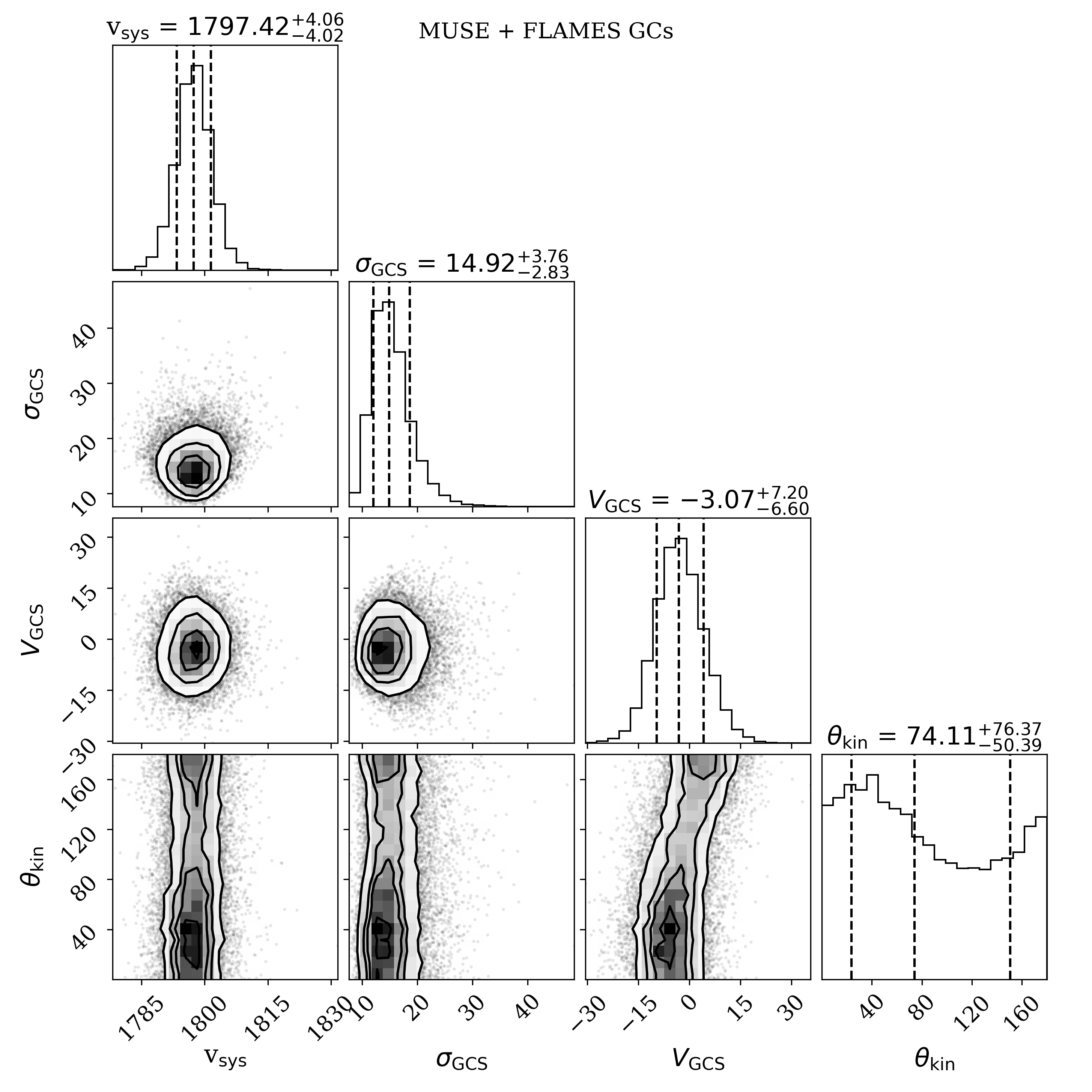}
    \caption{Corner plot \citep{corner} showing the posterior distributions of our simple kinematic model. The 16th, 50th, and 84th percentiles are indicated in the histograms with dashed lines. The kinematic angle is ill constrained due to the low velocity amplitude.}
    \label{fig:corner}
\end{figure}

Figure \ref{fig:corner} shows the posterior distributions of the fit. The parameters are well constrained, except for the position angle due to the low amplitude of rotation. We find a systemic velocity of $v_\text{sys} = 1797.45^{+4.02}_{-4.02}$ km s$^{-1}$, a rotation amplitude in agreement with zero, and a GCS velocity dispersion of $\sigma_\text{GCS} = 14.86^{+3.89}_{-2.83}$ km s$^{-1}$. Testing a model without rotation as described, for example in \cite{Emsellem2019}, we find similar values for the velocity dispersion.  
This value is higher than what was reported originally by \cite{vanDokkum2018_DM}, who found an intrinsic velocity dispersion of $\sigma_\text{int} = 3.2^{+5.5}_{-3.2}$ km s$^{-1}$ that was later revised to $\sigma_\text{int} = 5.6^{+5.2}_{-3.8}$ km s$^{-1}$ based on a revised velocity of GC98 \citep{vanDokkum2018_GC98_revision}. Also, the absence of rotation is in agreement with the measurements from KCWI observations of the stars presented by \cite{Danieli2019}. Using the data from \cite{vanDokkum2018_DM} and \cite{vanDokkum2018_GC98_revision}, \cite{Martin2018} used a kinematic model similar to the one presented here but that also allows for contamination and found $\sigma_\text{int} = 9.2^{+4.8}_{-3.6}$ km s$^{-1}$.

With the larger sample of GCs, we do not find any indication of a rotating GCS as proposed by \cite{Lewis2020_GCrotation} who used the velocities from \cite{vanDokkum2018_DM} and \cite{vanDokkum2018_GC98_revision} in a very similar model as the one used here. However, when using their GC sample, we found comparable values for the rotation and dispersion as in \cite{Lewis2020_GCrotation}, illustrating the effect small GC numbers combined with large velocity uncertainties can have. 
To test this further, we also ran the model using restricted GC samples containing only the MUSE GCs or only the FLAMES GCs, finding similar results. To test the influence of outliers (see \citealt{Laporte2019}), we randomly removed one or two GCs. However, the only GC that has a notable effect on the inferred velocity dispersion is GC101. Removing it, we find $\sigma=8.63^{+2.88}_{-2.14}$ km s$^{-1}$. As NGC\,1052-DF2 already has a relative radial velocity of $+315$ km s$^{-1}$ compared to the NGC\,1052 group \citep{vanDokkum2022_bullet}, it is unlikely that this GC is an interloper from the group. However, it is possible that this GC is in the process of being stripped from the galaxy or that its measured velocity is not reliable. In our testing, we consistently measured its elevated velocity, but given the rather low S/N, it is possible that the fit converges on a noise feature and deeper data would be required to exclude this possibility (see Appendix \ref{appendix}).

In summary, we find here a low velocity dispersion of the GCS that is comparable with previous estimates under consideration of the uncertainties. 
Also, the low dispersion of the GCs is comparable to the stellar velocity dispersion found from MUSE ($\sigma_\text{stars} = 10.8^{+3.2}_{-4.0}$ km s$^{-1}$; \citealt{Emsellem2019}) or KCWI ($\sigma_\text{stars} = 8.5^{+2.3}_{-3.1}$ km s$^{-1}$; \citealt{Danieli2019}). 

\section{Discussion}
\label{sect:discussion}
In this work, we reanalysed the GCS in NGC\,1052-DF2 using data from archival MUSE, with a focus on providing an updated view of their stellar populations and the GCS kinematics. In the following, we discuss our results in the context of previous studies.

\subsection{Dynamical mass estimates}
\label{sect:mass_estimate}
Connected to their measurements of the GCS velocity dispersion, several studies have estimated the dynamical or halo mass of NGC\,1052-DF2 using different mass estimators of varying complexity. In the original work, \cite{vanDokkum2018_DM} used a tracer mass estimator that considers the individual GC velocities as well as their spatial distribution in a potential. For different assumptions of the velocity anisotropy and at a distance of 20 Mpc, they infer an upper limited of the total mass within 7.6 kpc of $M_\text{dyn} < 3.4 \times 10^8 M_\sun$, while the stellar mass is estimated to be $M_\ast \sim 2 \times 10^8 M_\sun$ (from $L_V = 1.1 \times 10^8 M_\sun$). This implied a dynamical mass-to-light ratio ($M_\text{dyn}/L_V$) of order unity, therefore suggesting a lack of DM in this galaxy. Using a different approach to model the individual GC velocities from \cite{vanDokkum2018_DM}, \cite{Martin2018} found $\sigma = 9.5^{+4.8}_{-3.9}$ km s$^{-1}$ and $M_\text{dyn} < 3.7 \times 10^8 M_\sun$ within the effective radius (2.2 kpc = 22.7\arcsec; \citealt{vanDokkum2018_DM}), corresponding to $M_\text{dyn}/L_V$ < 6.7. 

Also incorporating the velocity dispersion of the stars, \cite{Emsellem2019} derived dynamical mass estimates from different estimators that are proportional to $R_e\,\sigma_\text{GCS}^2 G^{-1}$ \citep{Wolf2010, Amorisco2011, Courteau2014, Errani2018}. They found $M_\text{dyn}/L_V$ between $\sim$ 3.5 and 4.4 when assuming a distance of 20 Mpc, or $M_\text{dyn}/L_V$ between 5.8 and 6.0 when assuming a distance of 13 Mpc ($R_e = 1.4$\,kpc). In their rotation model, \cite{Lewis2020_GCrotation} find values of 
$M_\text{dyn}/L_V$ > 10 depending on the inclination of the system.

With a larger number of GCs and improved uncertainties, we provide here an updated dynamical mass estimate. Similar to \cite{Emsellem2019}, we employed a simple estimator for the mass within 1.8 of the effective radius $M_\text{dyn} (<1.8 R_e) = 6.3\,R_e\,\sigma_\text{GCS}^2 G^{-1}$ \citep{Errani2018}.
Assuming a distance of 16.2 Mpc ($R_e$ = 1.71 kpc), we find $M_\text{dyn} (<1.8\,R_e) = 5.6^{+3.3}_{-1.9} \times 10^8 M_\sun$ and $M_\text{dyn}/L_V = 8.7^{+5.1}_{-3.0}$ for $\sigma_{\text{GCS}}$ = 14.86 km s$^{-1}$. At a larger distance of 20.0 Mpc, this would correspond to $M_\text{dyn} (<1.8\,R_e) = 6.9^{+4.0}_{-2.4} \times 10^8 M_\sun$ and $M_\text{dyn}/L_V = 7.1^{+4.2}_{-2.5}$. Consequently, the higher velocity dispersion found here in combination with a smaller distance (Paper I), leads to a higher $M/L$ and a higher dynamical mass than originally proposed.

However, in the case GC101 truly is an interloper or being stripped, $\sigma$ is reduced to 8.63 km s$^{-1}$ and consequently $M_\text{dyn} (<1.8\,R_e) = 1.9^{+1.5}_{-0.8} \times 10^8 M_\sun$ and $M_\text{dyn}/L_V = 2.9^{+2.3}_{-1.3}$. At the larger distance of 20 Mpc, the mass-to-light ratio would further decrease to $M_\text{dyn}/L_V = 2.3^{+1.8}_{-1.3}$. In this case, the mass-to-light ratio of the galaxy is low, irrespective of the distance. 

\begin{figure}
    \centering
    \includegraphics[width=0.47\textwidth]{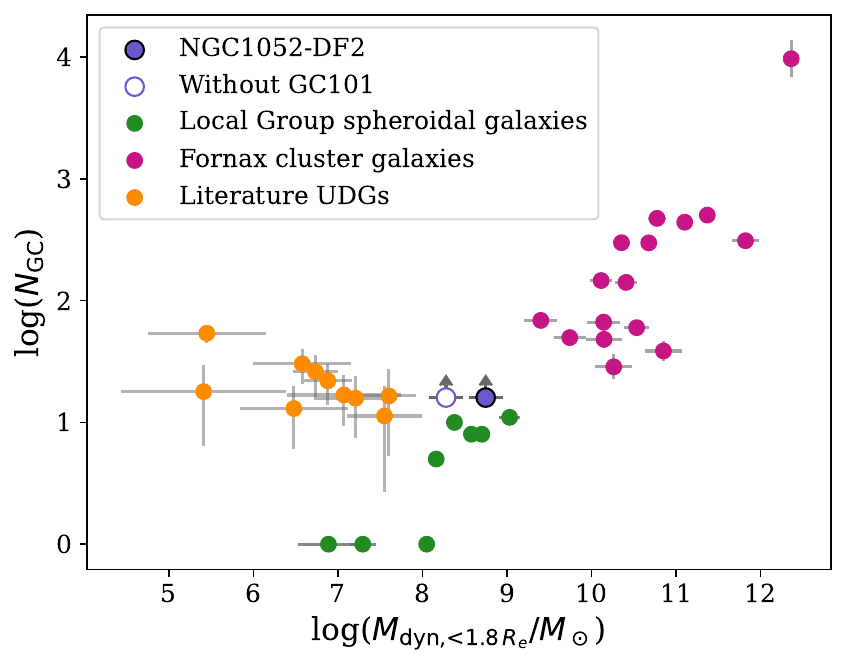}
    \caption{Total number of GCs versus the dynamical mass estimate derived from the GCS velocity dispersion and the galaxy effective radius. NGC\,1052-DF2 is shown in purple, Local Group dwarfs in green (data from \citealt{Forbes2018}), Fornax cluster galaxies in pink (data from \citealt{Fahrion2020b} and \citealt{Liu2019}), and literature UDGs in orange from the compilation by \citealt{Gannon2024}. The empty purple circle shows the dynamical mass estimate if GC101 is considered an interloper not belonging to the galaxy.}
    \label{fig:Mdyn_comparison}
\end{figure}

The dynamical mass of a galaxy is known to follow a tight relation with the total number of GCs or the total GCS mass (e.g. \citealt{SpitlerForbes2009, Harris2017, Forbes2018}), suggesting that both the galaxy halo and the GCS were established during early stages of galaxy formation. With its remarkably bright GCs and low velocity dispersion, NGC\,1052-DF2 was proposed to be an outlier of this relation (e.g. \citealt{vanDokkum2018_GCs}). To test this, we place NGC\,1052-DF2 on a plane showing dynamical masses versus GC numbers in Fig. \ref{fig:Mdyn_comparison}. As comparison samples, we used the Local Group spheroidal dwarf galaxies from \cite{Forbes2018} and calculated their dynamical masses with the simple mass estimator from \cite{Errani2018}. As a sample of more massive galaxies, we used galaxies in the Fornax cluster for which GCS velocity dispersions are available from \cite{Fahrion2020b} and GC numbers from \cite{Liu2019} and calculated their dynamical masses in the same fashion. Additionally, we used GC velocity dispersions and GC numbers complied by \cite{Gannon2024} for literature UDGs (in this case using data from \citealt{Beasley2016}, \citealt{Toloba2018}, \citealt{Forbes2019}, \citealt{Lim2020}, \citealt{Gannon2020}, \citealt{Gannon2021}, \citealt{Forbes2021}, \citealt{Danieli2022}, and \citealt{Toloba2023}). \cite{Gannon2024} also included NGC\,1052-DF2 in their catalogue, but we used here our updated measurements.

As can be seen from this figure, the updated dynamical mass estimate places NGC\,1052-DF2 among other similar galaxies (see also \citealt{Trujillo2019}), even with considering GC101 as an outlier. In particular, compared to other UDGs, NGC\,1052-DF2 has a rather large dynamical mass as estimated from its GCs. Nonetheless, we note here again that the reported dynamical mass is just meant as a simple estimate based on the velocity dispersion of the GCS and the host galaxy effective radius. To infer from this the halo mass, the DM profile would need to be modelled (see e.g. \citealt{Aditya2024}).

\subsection{Globular cluster metallicities}

The properties of the GCs in NGC\,1052-DF2 have been discussed extensively, in particular with respect to their extreme luminosities and large sizes when a distance of $D \sim$ 20 Mpc is assumed \citep{vanDokkum2018_GCs, Shen2021_GCLF}.
Besides their bright magnitudes, \cite{vanDokkum2022_GCs} found with an analysis of very deep HST ACS data that the bright GCs have a narrow colour range with a scatter of only $\sim$ 0.015 mag. The fainter (not spectroscopically confirmed) GCs analysed in their study appear to have a larger colour variation, but also larger uncertainties. Nevertheless, at first glance, the monochromatic nature of the GCs is in contrast to our finding of both metal-rich and metal-poor GCs. However, the GCs analysed in \cite{vanDokkum2022_GCs} were from the spectroscopically confirmed GCs known at the time, while four of the five metal-rich GCs found here were not covered in that analysis. Additionally, as shown in Fig. \ref{fig:CZR}, the colours reported by \cite{vanDokkum2022_GCs} fall into the region where the CZR has a steep slope. 
Consequently, a narrow colour range does not imply a narrow range of metallicities. 

While a broad distribution of GC metallicities is typically found in massive galaxies (e.g. \citealt{Peng2006}), dwarf galaxies in the Local Group are also known to host a rather large number of GCs with varying metallicities (see the compilation in \citealt{Beasley2019}). For example, NGC\,185 ($M_V$ = $-$14.8 mag; \citealt{McConnachie2012})
hosts GCs with metallicities between [M/H] = $-$2.0 and $-$0.5 dex, while NGC\,205 ($M_V = -14.6$ dex) has GCs with metallicities from [M/H] = $-1.2$ to $-0.5$ dex \citep{Sharina2006}. Both galaxies have similar luminosities as NGC\,1052-DF2 ($M_V$ = $-$14.9 mag from \citealt{vanDokkum2018_DM} at a distance of 16.2 Mpc) and also a substantial number of GCs, but have smaller effective radii \citep{McConnachie2012}.

\subsection{Implications for the formation of NGC\,1052-DF2}
Given the initial observed low DM fraction and luminous GCS, several mechanisms were discussed to explain NGC\,1052-DF2's properties and formation history. For example, an origin as an old tidal dwarf galaxy that has formed from gas expelled from a massive galaxy undergoing tidal interactions (e.g. \citealt{Mirabel1992, BournaudDuc2006, Ploeckinger2018, Haslbauer2019}) was discussed (e.g. \citealt{vanDokkum2018_DM, Fensch2019}). Assuming the closer distance of 16.2 Mpc combined with the our fiducial kinematic model, NGC\,1052-DF2 does not appear to lack DM, as would be expected for a tidal dwarf galaxy. With GC101 as an outlier, the dynamical mass estimate is lower, but even then the old stellar age and low metallicity of NGC\,1052-DF2 are at odds with the typically metal-rich and young populations found in tidal dwarfs (e.g. \citealt{Weilbacher2003, Duc2014}). As discussed by \cite{vanDokkum2018_DM} and \cite{Fensch2019}, the old age and low metallicity would imply a formation at high redshift. 

As an alternative scenario, tidal interactions with galaxies of the NGC\,1052-DF2 group were suggested to have stripped the DM halo of NGC\,1052-DF2 (e.g. \citealt{Maccio2021}). However, the closer distance would place NGC\,1052-DF2 in front of the NGC\,1052 group. Additionally, recent deep imaging of NGC\,1052-DF2 has shown no evidence of tidal interactions down to $\mu_g = 30.9$ mag arcsec$^{-2}$ \citep{Golini2024}. In contrast, the same study found tidal tails around NGC\,1052-DF4, another galaxy with a very low GCS velocity dispersion that has also been proposed to lack DM \citep{vanDokkum2019_DF4}. 
Nonetheless, as any stripping will always affect the DM first (e.g. see the discussions in \citealt{Montes2020}, \citealt{Smith2015}, and \citealt{Smith2016}), the low mass-to-light ratio found in the case GC101 is considered as an interloper together with the otherwise ordinary GC and stellar populations of NGC\,1052-DF2 could be an indication of an incomplete stripping. Additionally, the higher velocity of GC101 might be explained if this GC is in the process of being stripped.

To explain the properties and spatial configuration of both NGC\,1052-DF2 and NGC\,1052-DF4, \cite{vanDokkum2022_bullet} proposed a `bullet dwarf' collision scenario, in which the two UDGs were formed as a result of two gas-rich progenitor galaxies colliding (see also \citealt{Silk2019, Lee2024}). Reminiscent of the bullet galaxy cluster (e.g. \citealt{Clowe2004, Clowe2006}), the DM haloes get separated from the gas in the collision. As one consequence of this scenario, GCs are expected to form in a single event from the available gas and therefore should all have a single metallicity. In contrast, our finding of a range of GC metallicities instead implies different episodes of GC formation, through either continuous formation and enrichment of GCs at high redshifts or assembly through the mergers of subsystems as discussed for more massive galaxies (e.g. \citealt{Zepf1993, Beasley2020}). For this reason, if NGC\,1052-DF2 formed in such a scenario, at least some of the GCs were either already present or formed after the event (see also the discussion in \citealt{vanDokkum2022_GCs}). Also, it is unclear if a distance of 16.2 Mpc can be reconciled with this scenario.

\cite{TrujilloGomez2021} proposed a major merger origin of NGC\,1052-DF2 to explain the formation of its luminous GCS. While a lower distance to the galaxy results in a more ordinary GC mass function (see Fig. \ref{fig:masses}), a gas-rich merger might still have triggered star and cluster formation and could explain the range of GC metallicities if the some GCs were accreted in the merger while others were newly formed. The old GC ages place such a merger at high redshift, but the age uncertainties are unfortunately too large to constrain which GCs might have formed in such a merger.

If placed at 16.2 Mpc, many properties of NGC\,1052-DF2 are in line with Local Group dwarf galaxies, including its dynamical mass, GC number, GC mass function and the range of GC metallicities. However, it remains more extended than typical dwarf galaxies and still falls into the UDG category. Considering GC101 as an outlier, its dynamical mass also becomes rather low.
While our study does not allow us to exactly determine the formation history of NGC\,1052-DF2, it is possible that this galaxy is simply located in the tail of the size distribution of dwarf galaxies, or perhaps its stellar body has expanded. In general, several mechanisms have been proposed to explain the large sizes of UDGs, including external effects such as tidal interactions or ram pressure stripping (e.g. \citealt{Carleton2019, Sales2020}), internal feedback (e.g. \citealt{DiCintio2017, Chan2018}), or formation in high-spin haloes (e.g. \citealt{AmoriscoLoeb2016, Liao2019}). Using simulations, \cite{Benavides2024} recently showed that flat metallicity gradients in UDGs like the one found in NGC\,1052-DF2 are not expected from gradual evolution in high-spin haloes, but might rather be an effect of outflows from star formation or stripping. Similarly, also the NIHAO simulations found flat (or negative) gradients in UDGs formed via supernovae feedback \citep{CardonaBarrero2023}. Consequently, the flat stellar population gradients in NGC\,1052-DF2 together with the GC metallicity distribution might suggest that this galaxy has started out as a rather regular galaxy that has expanded due to feedback from intense star formation. 

\section{Conclusions}
This paper presents an updated view of the GCS of NGC\,1052-DF2. Using archival MUSE IFS data, we extracted the spectra of 12 GC candidates. We established the membership of 11 of them, including 4 GCs that have not yet been confirmed. Together with newly acquired high-resolution FLAMES spectra of 10 GCs, the number of spectroscopically confirmed GCs of NGC\,1052-DF2 is increased to 16. Our results are summarised as follows:
\begin{itemize}
    \item Using full spectrum fitting, we determined the individual ages and metallicities of the GCs in the MUSE field of view. We find overall old ages (> 9 Gyr) and a range of metallicities from [M/H] = $-0.73$ dex to [M/H] = $-1.86$ dex. Uncertainties on the GC metallicities range from $0.08$ to 0.24 dex.
    \item Comparing GC metallicities with their colours, we find them to scatter on the CZR. Nonetheless, with their low metallicities (and blue colours), they fall in a region where the slope of the CZR is steep and hence a small colour difference can correspond to a range of GC metallicities.
    \item We obtained radial profiles of the host stellar metallicity and ages, finding flat profiles within the uncertainties. Many of the GCs are more metal-poor than the host at the same position but have similarly old ages. 
    \item From the ages and metallicities, we then derived mass-to-light ratios using the predictions of the EMILES SSP models, finding values between 1.2 and 2.0 for the HST ACS F606W filter. By deriving photometric masses, we find that the GCS mass function peaks at $\sim 10^{5.4} M_\sun$, which is in agreement with the canonical peak of the GC mass function and the Milky Way GC mass function.
    \item We combined the GC velocities from MUSE with the high-precision velocities from FLAMES to build a simple kinematic model of the GCS using 16 GCs. We find a velocity dispersion of $\sigma_\text{GCS} = 14.86^{+3.83}_{-2.83}$ km s$^{-1}$ and no indication of rotation in the GCS. The velocity dispersion is insensitive to outliers with the exception of GC101. Excluding this GC, which has a low S/N spectrum, the velocity dispersion is reduced to $\sigma_\text{GCS} = 8.86^{+3.83}_{-2.83}$ km s$^{-1}$. Even if deeper data confirm the higher relative velocity of GC101, it could still be an interloper from the group or could be in the process of being stripped.    
    \item From this dispersion, we estimate a dynamical mass of $M_\text{dyn} (<1.8\,R_e) = 5.6^{+3.3}_{-1.9} \times 10^8 M_\sun$ and $M/L_V = 8.7^{+5.1}_{-3.0}$, similar to other dwarf galaxies with such a mass. Assuming GC101 is an interloper, we find $M_\text{dyn} (<1.8\,R_e) = 1.9^{+1.5}_{-0.8} \times 10^8 M_\sun$ and $M/L_V = 2.9^{+2.3}_{-1.3}$. Comparing the dynamical mass estimates to similarly derived values for other galaxies, NGC\,1502-DF2 does not appear as an outlier.
\end{itemize}

\noindent We find  NGC\,1052-DF2 to be an extended, low-mass galaxy with a rich GCS with a range of metallicities and a low dynamical mass. Its exact origin cannot be determined based on currently available data, but it is possible that it is a `puffed up' dwarf galaxy with a likely complex formation history shaped by hierarchical assembly of smaller structures and internal feedback.

\begin{acknowledgements}
We thank the anonymous referee for a constructive report that has helped to polish this manuscript.
We thank Mireia Montes and Ignacio Trujillo for helpful comments. This project has received funding from the European Union’s Horizon Europe research and innovation programme under the Marie Sk\l{}odowska-Curie grant agreement No 101103830.
KF acknowledges support through the ESA research fellowship programme.
TJ acknowledges the MUNI Award in Science and Humanities MUNI/I/1762/2023.
This work made use of Astropy:\footnote{\url{http://www.astropy.org}} a community-developed core Python package and an ecosystem of tools and resources for astronomy \citep{astropy2013, astropy2018, astropy2022}. Based on observations collected at the European Southern Observatory under ESO programmes 2101.B-5008(A) and 110.23P4.001.
\end{acknowledgements}

\bibliographystyle{aa} 
\bibliography{references}

\appendix
\section{Fitting GC101}
\label{appendix}
\begin{figure}
    \centering
    \includegraphics[width=0.98\linewidth]{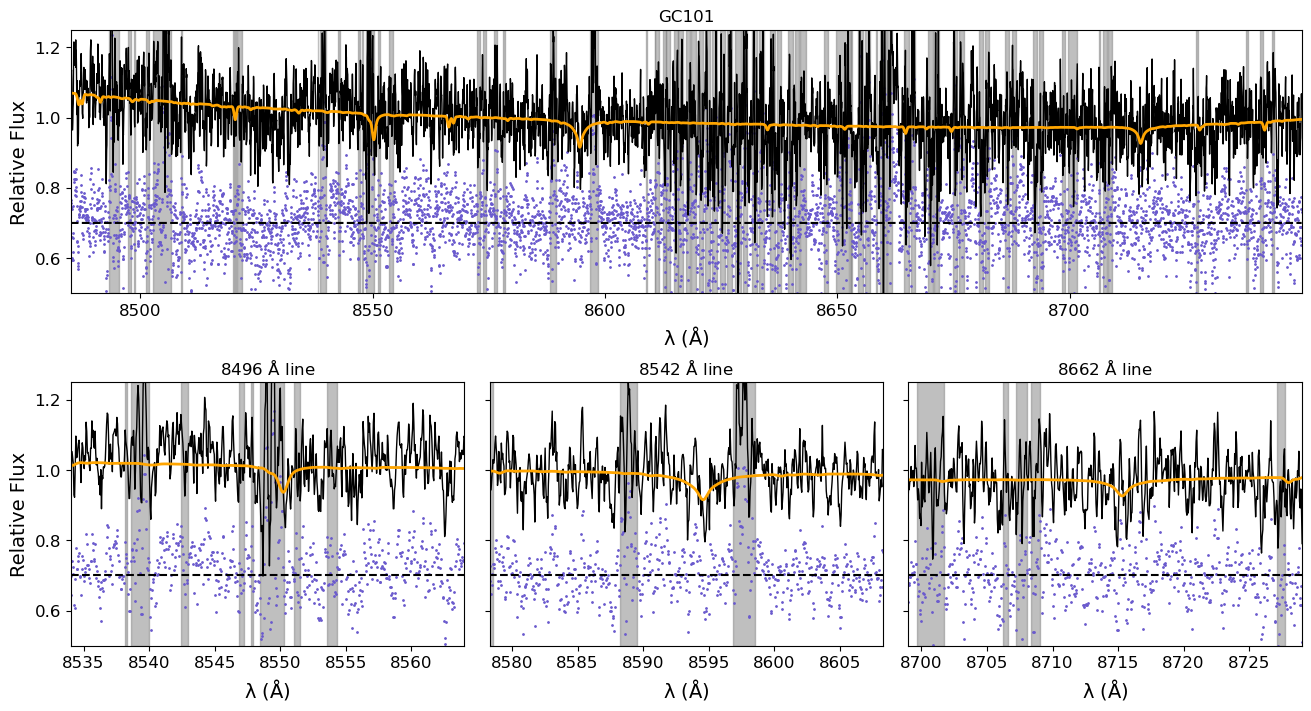}
    \caption{Full spectrum fit to the FLAMES spectrum of GC101. The original spectrum is shown in black, and the best-fit model is shown in orange. Shaded areas indicate masked regions due to sky line residuals. The lower panels show zoomed-in views around the calcium triplet lines.}
    \label{fig:GC101}
\end{figure}

We show the fit to the FLAMES spectrum of GC101 in Fig. \ref{fig:GC101}. The shown fit is using the stellar population models from \cite{Coelho2014}, similarly as was used to fit the other FLAMES spectra as described in Paper I. To test the robustness of the fit, we tested different starting velocities for the \textsc{pPXF} call, wavelength ranges and templates, consistently finding the elevated radial velocity.
However, as the figure shows, the spectrum is noisy with a S/N $\approx$ 5 pix$^{-1}$, whereas the other GCs observed with FLAMES have S/N between 6 and 19 pix$^{-1}$ (see Paper I). We can therefore not exclude systematic effects influencing the fit and deeper data would be needed to test this measurement.

\end{document}